\pdfoutput=1
\documentclass[nofootinbib,aps,prd,10pt]{revtex4}
\usepackage{silence}
\usepackage{amssymb,amsmath,graphicx,color,microtype}
\usepackage{enumerate}
\usepackage{slashed}
\usepackage{xcolor}
\usepackage{verbatim}
\usepackage{float}
\usepackage[utf8]{inputenc}
\graphicspath{{fig/}}

\usepackage{hyperref}
\hypersetup{
	colorlinks=true,
	linkcolor=blue,
	filecolor=blue,      
	urlcolor=blue, 
	citecolor=blue
}

\begin{document}

\title{A gigaparsec-scale local void and the Hubble tension} 

\author{Qianhang Ding}
\affiliation{Department of Physics, The Hong Kong University of Science and Technology, Hong Kong, P.R.China}
\affiliation{Jockey Club Institute for Advanced Study, The Hong Kong University of Science and Technology, Hong Kong, P.R. China}

\author{Tomohiro Nakama}
\affiliation{Department of Physics, The Hong Kong University of Science and Technology, Hong Kong, P.R.China}
\affiliation{Jockey Club Institute for Advanced Study, The Hong Kong University of Science and Technology, Hong Kong, P.R. China}

\author{Yi Wang}
\affiliation{Department of Physics, The Hong Kong University of Science and Technology, Hong Kong, P.R.China}
\affiliation{Jockey Club Institute for Advanced Study, The Hong Kong University of Science and Technology, Hong Kong, P.R. China}

\begin{abstract}
  We explore the possibility of using a gigaparsec-scale local void to reconcile the Hubble tension. Such a gigaparsec-scale void can be produced by multi-stream inflation where different parts of the observable universe follow different inflationary trajectories. The impact of such a void for cosmological observations is studied, especially those involving supernovae, Baryon Acoustic Oscillations (BAO) and the kinetic Sunyaev–Zel'dovich (kSZ) effect. As a benchmark model, a 1.7Gpc scale with boundary width 0.7Gpc and density contrast -0.14 may ease the Hubble tension.
\end{abstract}

\maketitle
\section{Introduction}

The value of the Hubble parameter is of central importance in modern cosmology. However, recent cosmological observations do not seem to converge on the measurement of the Hubble parameter.
Ref. \cite{Riess:2016jrr} reported a local value of the Hubble parameter $H_0=73.24\pm1.74 \mathrm{km}\,\mathrm{s}^{-1}\,\mathrm{Mpc}^{-1}$, $3.4\sigma$ higher than the value from Planck \cite{Aghanim:2016yuo}: $H_0=66.93\pm0.62 \mathrm{km}\,\mathrm{s}^{-1}\,\mathrm{Mpc}^{-1}$. Another independent determination by observations of multiply-imaged quasar systems yielded the results consistent with the local determination: $H_0=71.9^{+2.4}_{-3.0}\mathrm{km}\,\mathrm{s}^{-1}\,\mathrm{Mpc}^{-1}$ \cite{Bonvin:2016crt}. See also \cite{Feeney:2017sgx, Schoneberg:2019wmt, Lin:2019htv, DiValentino:2019qzk} for relevant discussions. 

While this so-called Hubble tension could be caused by unrecognized systematic uncertainties associated with the determinations of $H_0$ \cite{Freedman:2017yms, Rameez:2019wdt}, it has invoked studies of scenarios which can potentially alleviate or solve the tension by new physics. Such scenarios include early dark energy \cite{Karwal:2016vyq, Poulin:2018cxd, Alexander:2019rsc, Sakstein:2019fmf}, dark radiation  \cite{Riess:2016jrr, Bernal:2016gxb}, emerging spacial curvature \cite{Bolejko:2017fos}, evolving scalar fields \cite{Agrawal:2019lmo, Panpanich:2019fxq, Smith:2019ihp}, primordial non-Gaussianity \cite{Adhikari:2019fvb}, non-standard neutrino interactions (see \cite{Blinov:2019gcj} and references therein), acoustic dark energy \cite{Lin:2019qug}, neutrinos interacting with dark matter \cite{Ghosh:2019tab}, emergent dark energy \cite{Li:2019ypi}, quintessence axion dark energy \cite{Choi:2019jck}, a family of alternate dynamics of dark energy \cite{Mortsell:2018mfj}, massive dark vector fields \cite{Anchordoqui:2019yzc} and dissipative axion \cite{Berghaus:2019cls}, and so on. 

In addition, the Hubble tension could be explained a local void with radius $\simeq 150$ Mpc, but such a scenario was shown to be inconsistent with the supernova (SN) luminosity-distance relation (see \cite{Kenworthy:2019qwq} and references therein. See also \cite{Shanks:2018rka, Riess:2018kzi}). Here we explore a similar but different possibility, which employs a Gpc-scale local void. If we are located in the center of such a void, the Hubble tension could be alleviated, while evading the constraint of \cite{Kenworthy:2019qwq}. The presence of such a large void is unlikely in standard cosmological scenarios, as can be understood from $\sigma_8\simeq 0.81$ \cite{Aghanim:2018eyx}, which indicates that inhomogeneities with large amplitudes on comoving scales much larger than $10$ Mpc are statistically unlikely for Gaussian primordial fluctuations. However, such a large void can be realized in very early universe scenarios such as multi-stream inflation \cite{Li:2009sp}.

There is a long history about theories in which we live at the center of a Gpc-scale void. In particular, this possibility has been extensively discussed as an alternative to dark energy (see \cite{Biswas:2010xm, Li:2011sd} and references therein). Our proposal is also related to this line of research, but the depth of the void we need to ease or address the recently-debated Hubble tension is much smaller than that required for the void to be an alternative to dark energy. 

Ultimately, whether introducing a large void really helps or not in resolving the Hubble tension should be determined by testing such a hypothesis simultaneously against different observations of e.g. supernovae (SNe), baryon acoustic oscillations (BAO), large-scale structure and the cosmic microwave background (CMB), as was done in \cite{Biswas:2010xm} in the context of a void serving as an alternative to dark energy. However, we leave such a detailed analysis to a future work, partly because cosmological perturbation theory on an inhomogeneous background can be much more complicated than that for a homegeneous background. We instead, for simplicity and as a first step, just show how the local Hubble parameter behaves in a void cosmology, which would provide an indicator of how much the Hubble tension is eased as discussed later. In addition, a large void and hence its capabilities to ease the Hubble tension can be constrained by different observational effects of such a void, one of which is the kinetic Sunyaev-Zel'dovich (kSZ) effect \cite{Sunyaev:1980nv}, and we also show how a large void is strictly constrained by the kSZ effect.

\section{A large local void realized in multi-stream inflation}

A gigaparsec-scale void with $\mathcal{O}(0.1)$ density contrast is exponentially unlikely to arise from rare random fluctuations of a minimal scenario of cosmic inflation. To better motivate our study, we show that such a void can be generated by multi-stream inflation \cite{Li:2009sp}. In multi-stream inflation, the inflationary trajectory bifurcates in multi-field space by encountering a barrier or waterfall type potential. Previously multi-stream inflation was also used to generate position space features in our universe such as a cold spot on the CMB \cite{Afshordi:2010wn}, initial clustering of primordial black holes \cite{Ding:2019tjk} and structures in the multiverse \cite{Li:2009me}.

\begin{figure}[htbp] \centering
  \includegraphics[width=0.8\textwidth]{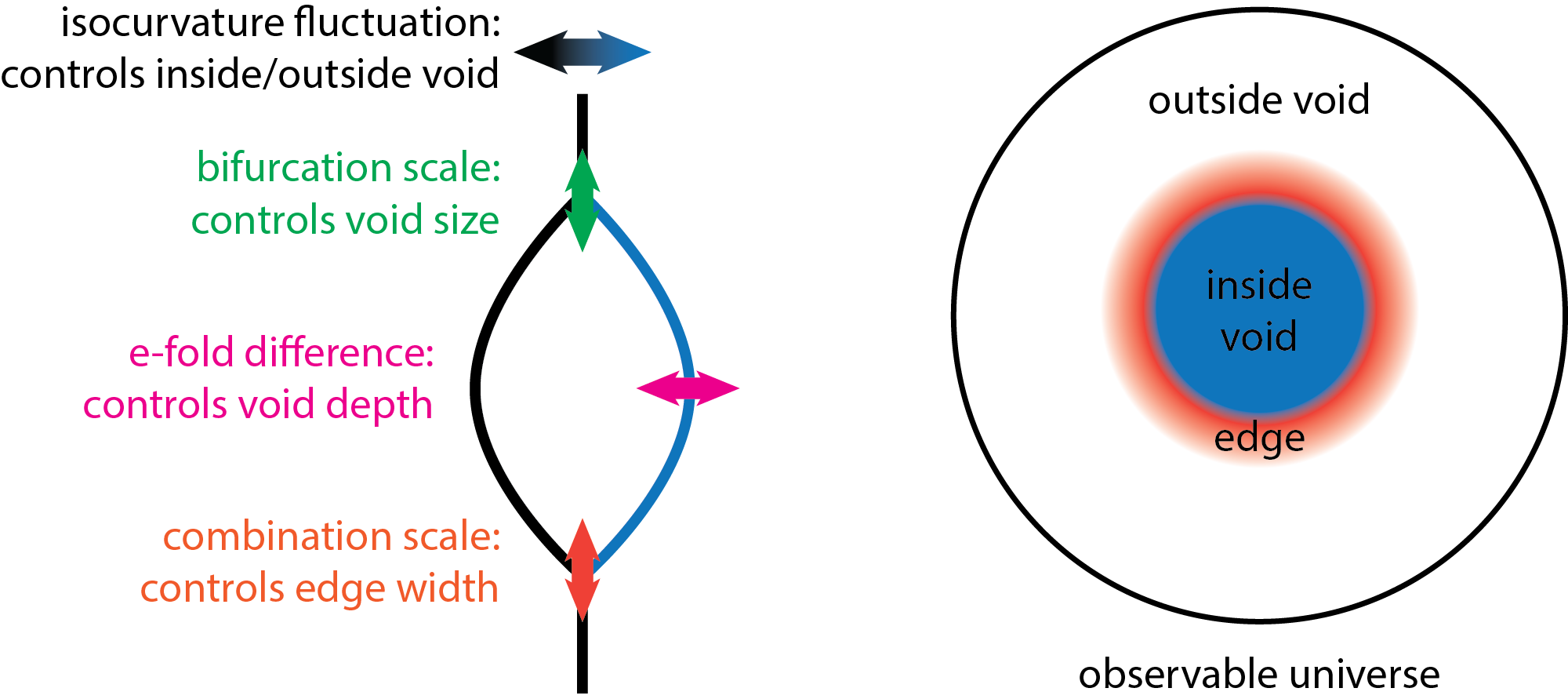}
  \caption{\label{fig:msi}
    A cosmic void generated from multi-stream inflation. 
    \emph{Left panel:} multi-stream inflation in the primordial universe. The inflationary trajectory (vertical direction from top to down) encounters a barrier or waterfall potential, which causes bifurcation of the inflationary trajectory. With a larger probability the inflaton rolls along the black trajectory on the left, but with a smaller probability it rolls along the blue trajectory on the right. The region experiencing this blue trajectory later shows up as a local void. 
    \emph{Right panel:} The late-time universe arises from multi-stream inflation. 
  }
\end{figure}

We illustrate the dynamics of multi-stream inflation in Fig.~\ref{fig:msi}. If the inflaton trajectory gets split during inflation, and further if the potential realizing such a situation is constructed in such a way that spatial regions corresponding to one of the trajectories experience shorter inflation than the other regions corresponding to the other trajectory, then such regions later show up as underdense regions or voids, where the matter energy density contrast is related to the e-folding number difference by $\delta\rho / \rho \sim \delta N$. The abundance and depth of such voids as well as the density profile for each void can be controlled by the potential shape realizing such a bifurcation of the inflaton trajectory. 

If the bifurcation probability is small, it is possible that only one such void exists in the observable universe. In this case, it is more likely that the void is of spherical shape. Because events with small chance, either rare tail of Gaussian fluctuations in the isocurvature fluctuation or quantum tunneling, prefer spherical bubbles.

The width of the void boundary is determined by the e-folding number between the bifurcation and combination of the trajectory, since the edge was decoupled from the Hubble flow during the bifurcated stage of inflation, and then expanded again after combination. To get a void whose boundary is thick, we require that the bifurcation last only for of order one e-fold.

\section{Void Cosmology}
\begin{figure}[htbp]
\centering
\includegraphics[width=8cm]{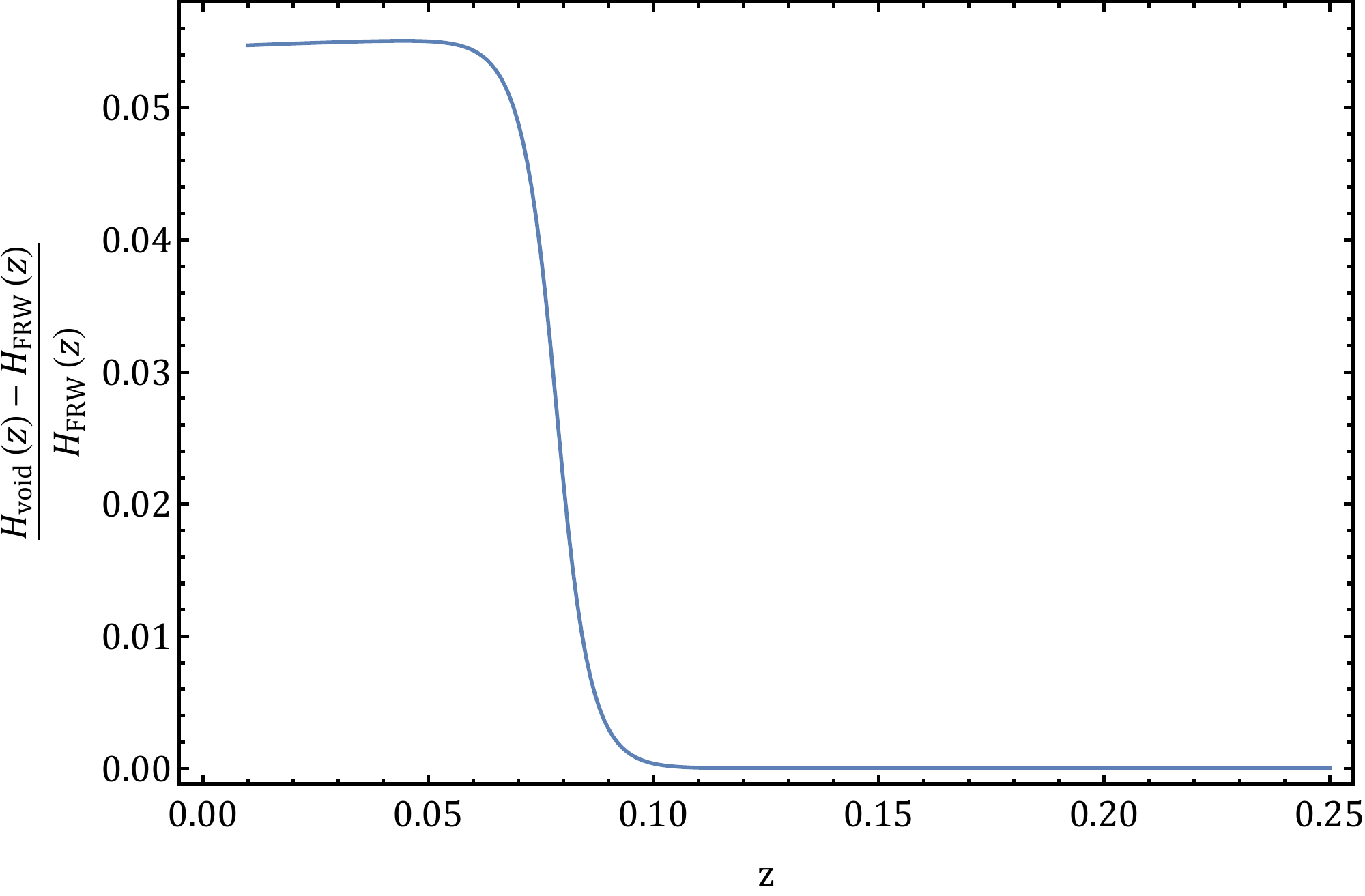} \hspace{0.05\textwidth}
\includegraphics[width=8cm]{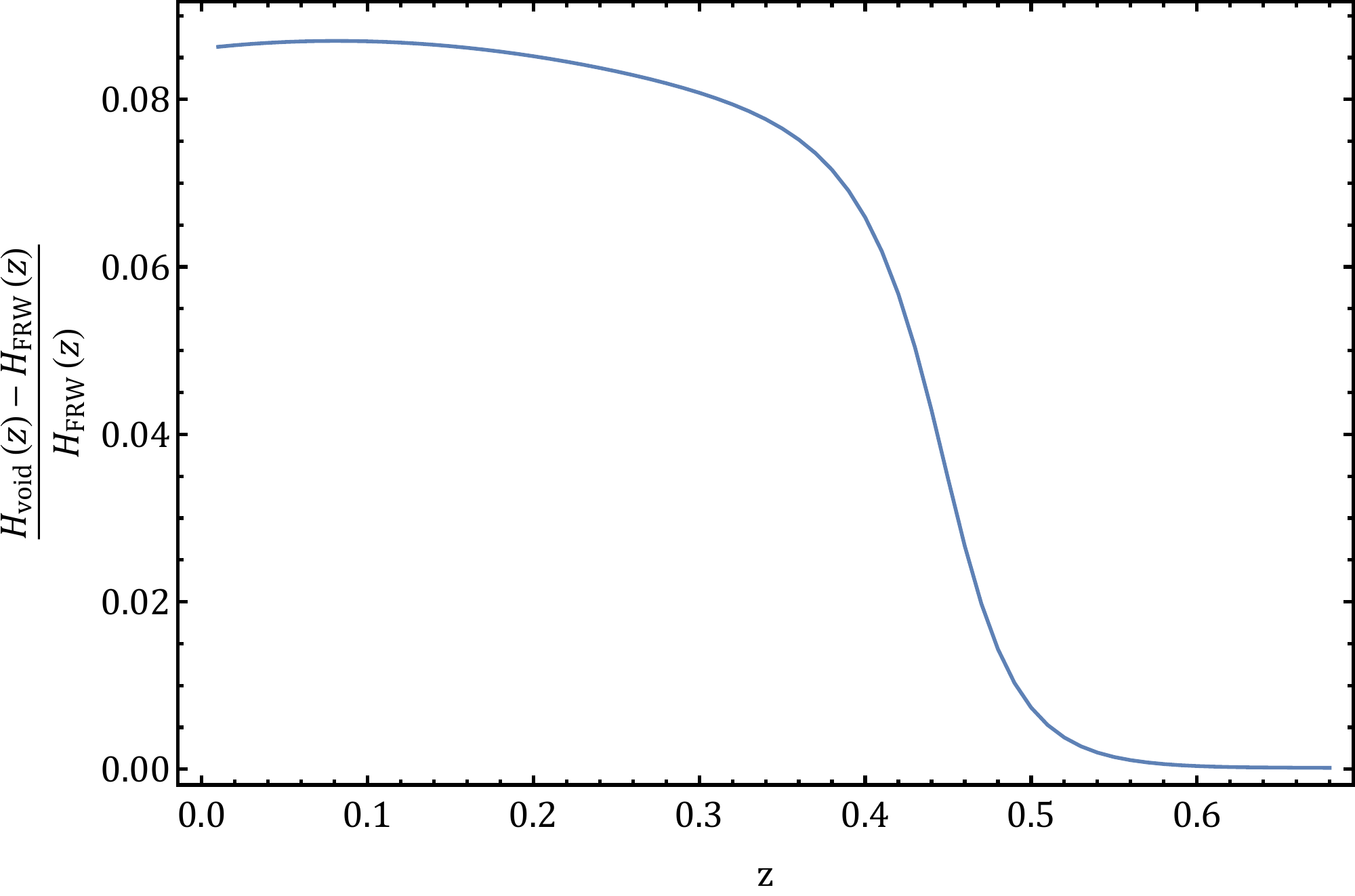}
\caption{The left panel shows the difference between $H_{\mathrm{void}}(z)$ and $H_{\mathrm{FRW}}(z)$ based on the KBC void parameters and $H_{\mathrm{0,out}} = 73.2 \mathrm{km}\,\mathrm{s}^{-1}\,\mathrm{Mpc}^{-1}$ \cite{Kenworthy:2019qwq}. The right panel shows the difference between $H_{\mathrm{void}}(z)$ and $H_{\mathrm{FRW}}(z)$ for $\delta_V = -0.4439$, $r_V = 1700 \mathrm{Mpc}$, $\Delta_r = 102 \mathrm{Mpc}$, $H_{\mathrm{0,out}} = 67.4 \mathrm{km}\,\mathrm{s}^{-1}\,\mathrm{Mpc}^{-1}$, $\Omega_M = 0.315$ and $\Omega_{\Lambda} = 0.685$.}
\label{correction}
\end{figure}

\begin{figure}[H]
\centering
\includegraphics[width=8cm]{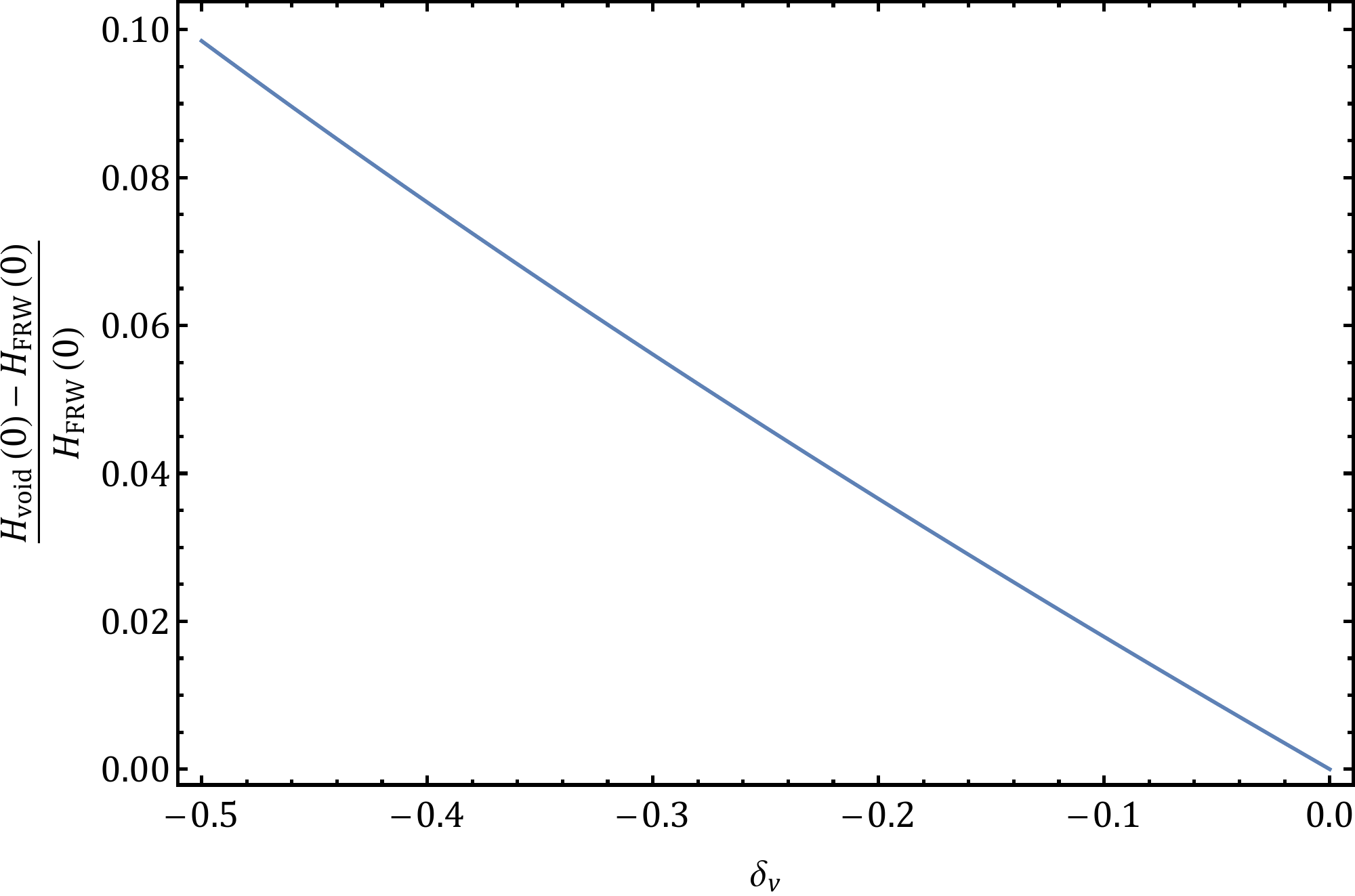}
\caption{The figure shows the void depth dependence of the difference between  $H_{\mathrm{void}}(0)$ and $H_{\mathrm{FRW}}(0)$.}
\label{hubbdep}
\end{figure}

\begin{figure}[H]
\centering
\includegraphics[width=12cm]{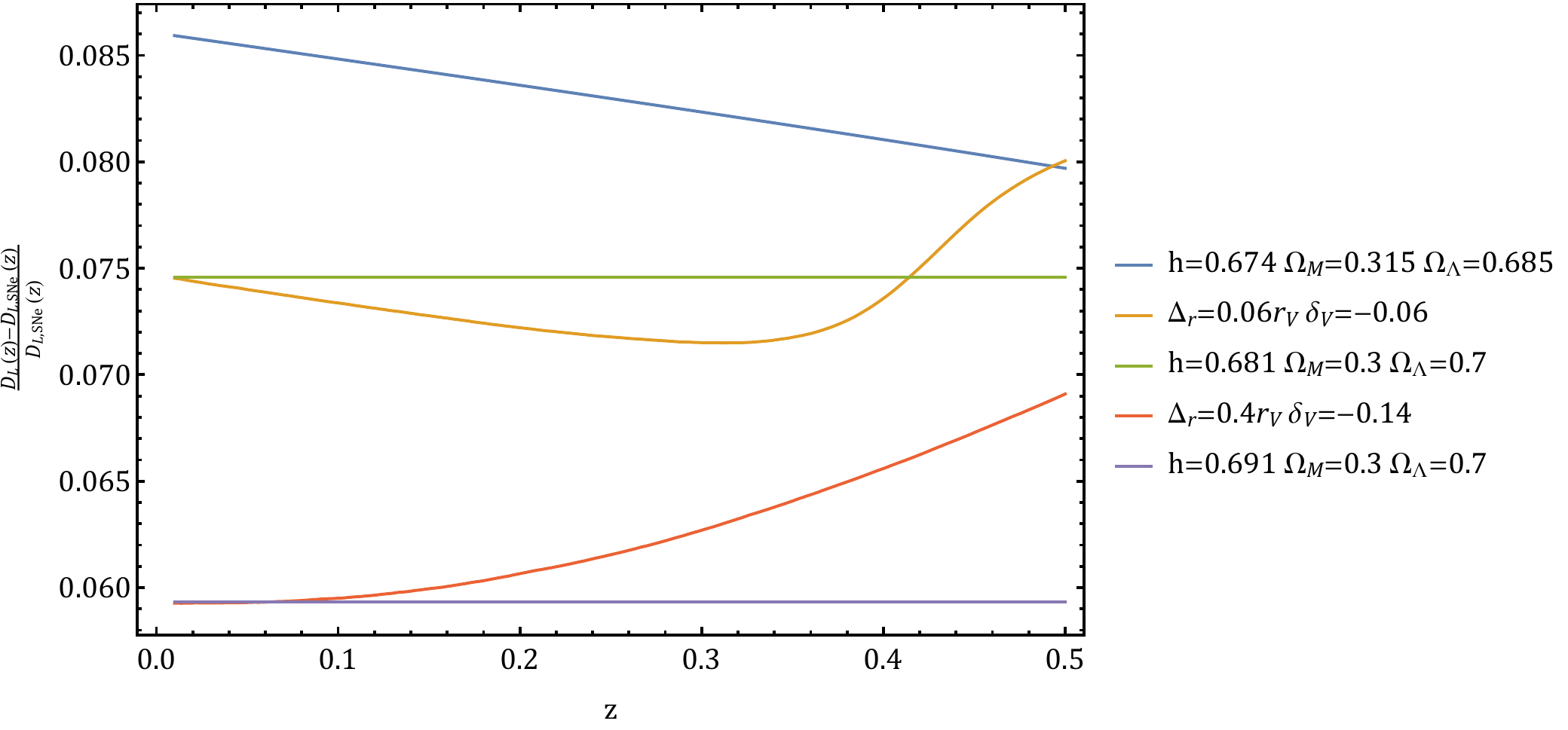}
\caption{The difference in the luminosity distance for different cosmologies from that ($D_{\mathrm{L,SNe}}(z)$) for an FRW universe with $(h,\Omega_M,\Omega_\Lambda)=(0.732,0.3,0.7)$, where $h=H_0/(100\mathrm{km/s/Mpc})$. Here as well as elsewhere the outer boundary conditions for the void cosmologies are given by Planck 2018 ($(h,\Omega_M,\Omega_\Lambda)=(0.674,0.315,0.685)$). The value $h=0.681 (0.691)$ corresponds to $H_{\mathrm{void}}(z=0)=68.1 (69.1)\mathrm{km/s/Mpc}$ for the void case with $(\Delta_r/r_V,\delta_V)=(0.06,-0.06) ((0.4,-0.14))$. \label{luminosity}}
\end{figure}

We follow \cite{Kenworthy:2019qwq} to describe a local void, where more details can be found. We use the following Lemaitre-Tolman-Bondi (LTB) metric \cite{Lema1997, tolman1934, bondi1947} (see also \cite{Kenworthy:2019qwq}):
\begin{align}
ds^2 = c^2 dt^2 - \frac{R'(r,t)^2}{1 - k(r)} dr^2 - R^2(r,t) d\Omega^2~,
\end{align}
where $R'(r,t) = \partial R(r,t)/\partial r$. For a homogeneous case, $R(r,t) = a(t) r$, $k(r) = k r^2$, where $a(t)$ is the usual scale factor. Then, we have the following Friedmann equation \cite{GarciaBellido:2008nz}:
\begin{align} \label{fredmann}
H(r,t)^2 \equiv \frac{\dot{R}(r,t)^2}{R(r,t)^2} = H_0(r)^2 \left(\Omega_M (r) \frac{R_0(r)^3}{R(r,t)^3}+\Omega_k (r) \frac{R_0(r)^2}{R(r,t)^2}+\Omega_\Lambda (r)\right)~.
\end{align}
Here, $H_0(r)\equiv H(r, t_0)$, $\Omega_M (r)\equiv\rho_M (r)/\rho_c(r)$ with $\rho_c(r)\equiv 3 H_0(r)^2/8 \pi G$, $\Omega_\Lambda(r)\equiv \rho_\Lambda/\rho_c (r)$ and $\Omega_k (r)\equiv -c^2 k(r)/H_0(r)^2 R_0(r)^2$. These quantities satisfy $\Omega_M(r) + \Omega_k(r) + \Omega_\Lambda(r) = 1$. Let us introduce
\begin{align}
\delta(r) \equiv \frac{\rho_M(r) - \rho_M(\infty)}{\rho_M(\infty)}~.
\end{align}
We assume the following form for $\delta(r)$ \cite{Kenworthy:2019qwq}:
\begin{align}\label{densitydis}
\delta(r) = \delta_V \frac{1 - \tanh((r - r_V)/2\Delta_r)}{1 + \tanh(r_V/2\Delta_r)}~.
\end{align}
Here, $\delta_V$ and $r_V$ represent the depth and radius of the void, and $\Delta_r$ represents the width of the void edge. In Kenworthy-Scolnic-Riess (KBC) \cite{Kenworthy:2019qwq}, the parameters of the void \cite{Hoscheit:2018nfl} are chosen as $\delta_V = -0.3$, $r_V = 308 \mathrm{Mpc}$, $\Delta_r = 18.46 \mathrm{Mpc}$. We will mainly be interested in a larger and shallower void, as will be specified later.
One can use Hubble constant outside the void to express the energy density as \cite{Kenworthy:2019qwq}
\begin{align}\label{density}
\rho_M(r) &\propto \Omega_M (r) H_0 (r)^2 = \Omega_{M,\mathrm{out}} (1+\delta (r)) H_{\mathrm{0,out}}^2~,\\\label{vacuum}
\rho_{\Lambda} (r) &\propto \Omega_{\Lambda} (r) H_0 (r)^2 = (1-\Omega_{M,\mathrm{out}}) H_{\mathrm{0,out}}^2 = \mathrm{const}~.\\
\rho_k (r) &\propto \Omega_{k} (r) H_0 (r)^2 = (1-\Omega_M (r)-\Omega_{\Lambda} (r)) H_0 (r)^2 = H_0 (r)^2 - (1 + \Omega_{M,\mathrm{out}} \delta (r)) H_{\mathrm{0,out}}^2~.
\end{align}
Here and hereafter the subscripts ``out'' denote quantities outside a void. Then we can express Eq.~\eqref{fredmann} into
\begin{align}\label{fredmann1}
\frac{\dot{R}(r,t)^2}{R(r,t)^2} = \Omega_{M,\mathrm{out}} (1+\delta (r)) H_{\mathrm{0,out}}^2 \frac{R_0(r)^3}{R(r,t)^3}+(H_0 (r)^2 - (1 + \Omega_{M,\mathrm{out}} \delta (r)) H_{\mathrm{0,out}}^2)\frac{R_0(r)^2}{R(r,t)^2} + (1-\Omega_{M,\mathrm{out}}) H_{\mathrm{0,out}}^2~.
\end{align}
We choose the gauge $R(r, t_0) = R_0(r) = r$ \cite{GarciaBellido:2008nz}. Then
\begin{align}\label{fredmann2}
\frac{\dot{R}(r,t)^2}{R(r,t)^2} = \Omega_{M,\mathrm{out}}(1+\delta (r)) H_{\mathrm{0,out}}^2  \frac{r^3}{R(r,t)^3}+(H_0 (r)^2 - (1 + \Omega_{M,\mathrm{out}} \delta (r)) H_{\mathrm{0,out}}^2)\frac{r^2}{R(r,t)^2} + (1-\Omega_{M,\mathrm{out}})  H_{\mathrm{0,out}}^2~.
\end{align}
Integrating Eq.~\eqref{fredmann2},
\begin{align}\label{intfredmann}
t_B(r) = \int_0^r dR~ R^{-1} \left[\Omega_M(r) H_0(r)^2 \left(\frac{r}{R}\right)^3 + \Omega_k(r) H_0(r)^2 \left(\frac{r}{R}\right)^2 + \Omega_\Lambda(r) H_0(r)^2\right]^{-1/2}~.
\end{align}
Here, $t_B$ is the time since big bang. As in \cite{Kenworthy:2019qwq},  we choose $t_B (r) = t_B = \mathrm{const}$. and identify the age of the universe $t_0=t_B$. When $r\gg r_V$, $\delta (r) = 0$ as well as $H_0 (r) = H_{\mathrm{0,out}}$, and hence we get
\begin{align}
t_B = \int_0^1 \frac{d a_{\mathrm{out}}}{H_{\mathrm{0,out}} [\Omega_{M,\mathrm{out}} a_{\mathrm{out}}^{-1} + \Omega_{\Lambda,\mathrm{out}} a_{\mathrm{out}}^2]^{1/2}}~.
\end{align}
Here, $a_{\mathrm{out}}$ is the usual scale factor outside a void. From this equation we can obtain $t_B$ by specifying $\Omega_{M,\mathrm{out}}$ and $\Omega_{\Lambda,\mathrm{out}}$, then for any given radius $r$, we can use Eq.~\eqref{intfredmann} to obtain $H_0 (r)$. Then, we can numerically calculate $R (r, t)$ by solving
\begin{align}\label{intfredmann1}
\frac{\partial R}{\partial t} = R \left[ H_0(r)^2 \Omega_M(r) \frac{r^3}{R^3} + H_0(r)^2 \Omega_k(r)\frac{r^2}{R^2} + H_0(r)^2 \Omega_{\Lambda}(r) \right]^{1/2}~.
\end{align}
We provide initial condition at $R(r,t_B)=r$ and solve this differential equation backward in time from $t_B$ to $t(<t_B)$ to obtain $R(r,t)(<r)$. 
To get $H_{\mathrm{void}}(z)\equiv H (r(z), t(z))$, we need $R (r(z), t(z))$ along the trajectory of photons reaching $r=0$ now.
From the null geodesics equation, we can numerically get the relation between $r$ and redshift $z$ and the relation between $t$ and redshift $z$ \cite{Kenworthy:2019qwq}:
\begin{align}\label{dtr}
\frac{dt}{dr} = -\frac{1}{c}\frac{R'(r,t)}{\sqrt{1-k(r)}}~,\\\label{dzr}
\frac{1}{1+z}\frac{dz}{dr} = \frac{1}{c}\frac{\dot{R}'(r,t)}{\sqrt{1-k(r)}}~.
\end{align}
The Hubble parameter in an FRW universe $H_{\mathrm{FRW}}(z)$ is written as 
\begin{align}
H_{\mathrm{FRW}}(z) = H_{\mathrm{0,out}} \sqrt{\Omega_M (1+z)^3 + \Omega_{\Lambda}}~.
\end{align}
Then, we calculate the difference between $H_{\mathrm{void}}(z)$ and $H_{\mathrm{FRW}}(z)$, which is shown in Fig.~\ref{correction}.
We also calculate the relative difference between $H_{\mathrm{void}}(z)$ and $H_{\mathrm{FRW}}(z)$ at redshift $z = 0$ as a function of $\delta_V$, which is shown in Fig.~\ref{hubbdep}.
This quantity is expected to provide an indicator of how much the Hubble tension between the $H_0$ determination involving SNe observations and the CMB $H_0$ determination is eased. This is because the luminosity distance as a function of redshift in a void cosmology, given by $(1+z)^2R(r(z),t(z))$ \cite{Kenworthy:2019qwq},  resembles that in a flat FRW cosmology with $H_0$ corresponding to $H_{\mathrm{void}}(z=0)$ in a void cosmology, as shown in Fig.~\ref{luminosity}. 
The behaviors of the curves can be understood from the Taylor expansion of the luminosity distance by redshift for an FRW universe, which looks like $D_L(z)=cz/H_0(1+\cdots)$ (see e.g. \cite{Kenworthy:2019qwq}). This shows that the luminosity distance at low $z$ is primarily determined by $H_0$, and $\Omega_m,\Omega_\Lambda,\Omega_k$, appearing from the second term, give relatively minor corrections. Similar understanding would hold for void cosmologies. Though a more careful analysis would be merited, introducing a local void is likely to ease the so-called Hubble tension, judging from this figure. 
 
\begin{figure}[htbp]
\centering
\includegraphics[width=8cm]{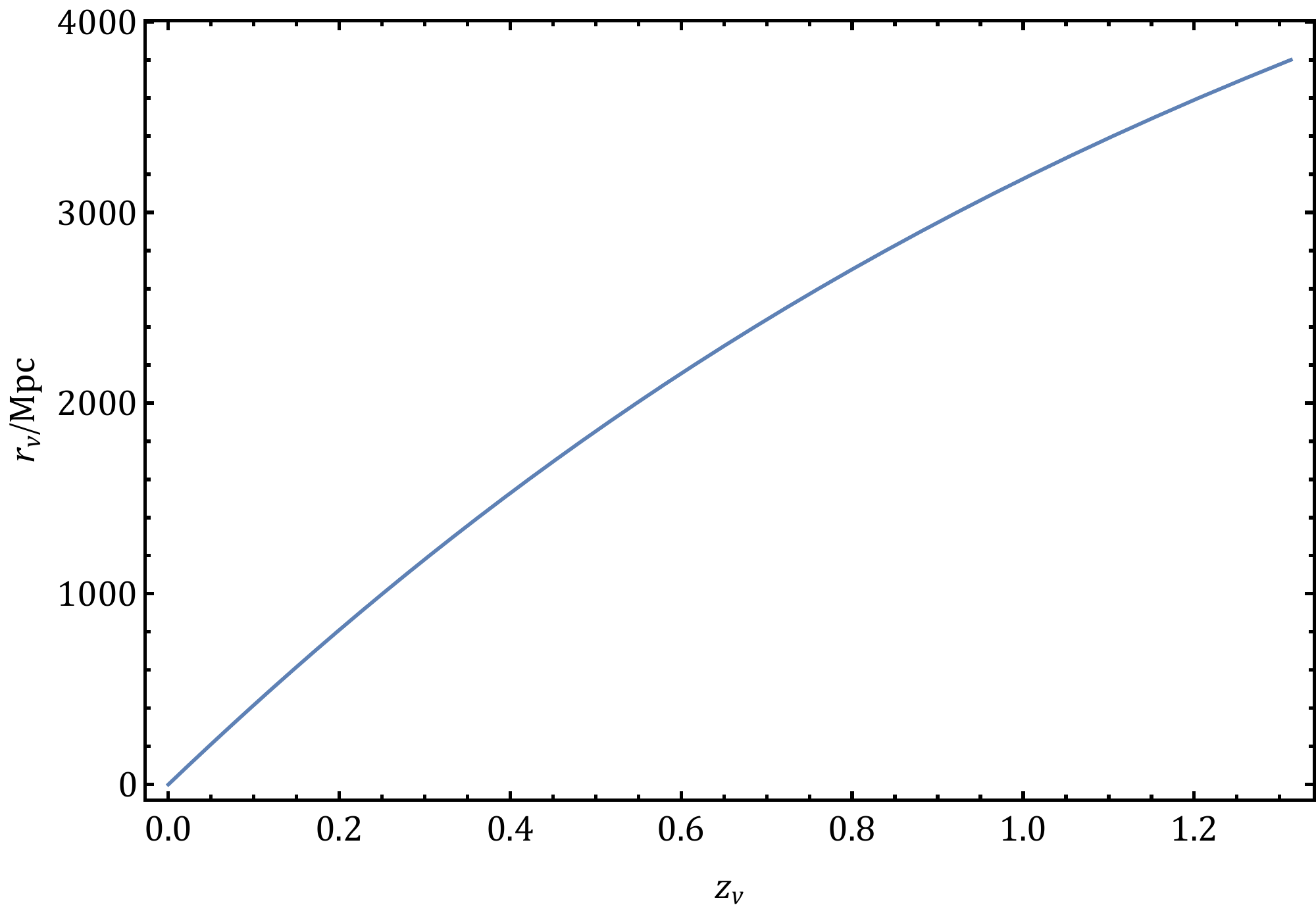}
\caption{The relation between the void size $r_V$ and the corresponding redshift $z_V$ for $\Delta_r = 0.06 r_V$.}
\label{rzv}
\end{figure}

From the Eqs. \eqref{dtr} and \eqref{dzr}, we can get the redshift $z_V$ corresponding to $r_V$, which is shown in Fig.~\ref{rzv}. For each $r_V$, we keep $H_{\mathrm{0,in}} = 73.2 \mathrm{km}\,\mathrm{s}^{-1}\,\mathrm{Mpc}^{-1}$ and $H_{\mathrm{0,out}} = 67.4 \mathrm{km}\,\mathrm{s}^{-1}\,\mathrm{Mpc}^{-1}$.

\section{BAO Observation}
Baryon Acoustic Oscillations (BAO) provide a standard ruler and probing the BAO scale at different redshift is useful in constraining cosmological models.
BAO signals are detected by observing galaxies with redshift $0.106\lesssim z \lesssim 0.61$ \cite{beutler2011, Ross:2014qpa, Alam:2016hwk}, the angular distance and the expansion rate are determined at $z=2.34$ by observing the Ly$\alpha$ forest of high-redshift quasars \cite{Delubac:2014aqe, Font-Ribera:2013wce, Bautista:2017zgn, Bourboux:2017cbm, Blomqvist:2019rah, Agathe:2019vsu} and the acoustic scale at $z\sim 1100$ is determined from CMB \cite{Aghanim:2018eyx}.

For the sound horizon $r_d$ at the drag epoch, an important quantity in analysing BAO data, there is a numerically calibrated approximate formula \cite{Aubourg:2014yra}:
\begin{align}
r_d \approx \frac{55.154 \exp[-72.3 (\omega_{\nu} + 0.0006)^2]}{\omega_M^{0.25351} \omega_B^{0.12807}}\mathrm{Mpc}~,\label{rd}
\end{align}
where $\omega_X = \Omega_X h^2$, with $X = M, \nu, B$ representing matter, neutrinos and baryons respectively, and $h = H_0/(100\mathrm{km/s/Mpc})$. 

In order to get the sound horizon, we may use the BBN determinations of the baryon density. From a theoretical estimation and observations \cite{Cooke:2017cwo}, 
\begin{align}
100 \Omega_B h^2 = 2.166 \pm 0.015 \pm 0.011 ~~~~~\mbox{(BBN theoretical)},\\
100 \Omega_B h^2 = 2.235 \pm 0.016 \pm 0.033 ~~~~~\mbox{(BBN empirical)}~.
\end{align}
The Planck 2018 \cite{Aghanim:2018eyx} also provides the baryon density
\begin{align}
100 \Omega_B h^2 = 2.237 \pm 0.015 ~~~~~\mbox{(Planck 2018)}~. \label{omegab}
\end{align}
Then we can use BAO data to determine $\Omega_m$ and $H_0$ for FRW cosmologies. Such a result is shown in Fig.~2 of \cite{Cuceu:2019for}, where $\Lambda$CDM is assumed. For each galaxy BAO and Ly$\alpha$ BAO, its center value is close to the Hubble constant from SNe. However, their intersection indicates values close to the Hubble constant from the Plank observations.

We try to illustrate implications of a local large void for BAO interpretations as follows. Ref. \cite{Biswas:2010xm} discussed how a model-independent physical observable $(\Delta\theta^2\Delta z)^{1/3}$ at different redshift $z_{\mathrm{BAO}}$ for BAO observations is calculated for FRW and void cosmology. This quantity is related to the sound horizon at the drag epoch $r_d$. It was noted in \cite{Aubourg:2014yra} that there are different conventions for $r_d$, which can differ 1-2\% level. The above formula (Eq.~(\ref{rd})) is for the CAMB convention. Note that it doesn't depend on $\Omega_k$, since it is determined by micro-physics at the drag epoch.   

Equations needed to calculate $(\Delta \theta^2\Delta z)^{1/3}$ for FRW cases can be found in \cite{Biswas:2010xm}:
\begin{equation}
    (\Delta\theta^2\Delta z)^{1/3}=\frac{z_{\mathrm{BAO}}^{1/3}r_d}{D_V^{\mathrm{FRW}}(z_{\mathrm{BAO}})},
\end{equation}
where for flat FRW cases
\begin{equation}
    D_V^{\mathrm{FRW}}(z_{\mathrm{BAO}})=\frac{1}{H_0}\left[\frac{z_{\mathrm{BAO}}}{h(z_{\mathrm{BAO}})}\left(\int_0^{z_{\mathrm{BAO}}}\frac{dz}{h(z)}\right)^2\right]^{1/3}
\end{equation}
with $h(z)=H(z)/H_0$. 

We calculate $(\Delta \theta^2\Delta z)^{1/3}$ for void cosmologies similarly to \cite{Biswas:2010xm} as follows. For void cosmology $r_d$ depends on $r$, hence we estimate $r_d(r)$ simply by replacing in Eq.\,(\ref{rd}) 
 $\omega_{(M,B)}\rightarrow \omega_{(M,B)}(r)=\Omega_{(M,B),\mathrm{out}}(1+\delta(r))h_{0,\mathrm{out}}^2$ with $h_{0,\mathrm{out}}=H_{0,\mathrm{out}}/(100\mathrm{km/s/Mpc})$ (see Eq.~(\ref{density})). For neutrinos we use $\omega_\nu=\Sigma m_\nu/93.14$eV \cite{Lesgourgues:2012uu} assuming $\Sigma m_\nu=0.06$eV \cite{Aghanim:2018eyx}. We also need the $r$-dependent drag epoch $z_d(r)$, for which we use the following modified versions of the formulae from \cite{Eisenstein:1997ik}:
 \begin{equation}
     z_d(r)=\frac{1291\omega_M(r)^{0.251}}{1+0.659\omega_M(r)^{0.828}}(1+b_1(r)\omega_B(r)^{b_2(r)}),
 \end{equation}
 \begin{equation}
     b_1(r)=0.313\omega_M(r)^{-0.419}(1+0.607\omega_M(r)^{0.674}),
 \end{equation}
 \begin{equation}
     b_2(r)=0.238\omega_M(r)^{0.223}.
 \end{equation}
Let us introduce the radius $r_{\mathrm{BAO}}$ and time $t_{\mathrm{BAO}}$, which are the radius and time corresponding to $z_{\mathrm{BAO}}$, where the correspondence is determined by solving the geodesic equation from the observer at $(t_0,r=0)$. We further introduce $R_{\mathrm{BAO}}=R(r_{\mathrm{BAO}},t_{\mathrm{BAO}})$. We determine the drag time $t_d(r_{\mathrm{BAO}})$ by 
\begin{equation}
    \frac{1+z_d(r_{\mathrm{BAO}})}{1+z_{\mathrm{BAO}}}=\left(\frac{R'_{\mathrm{BAO}}R^2_{\mathrm{BAO}}}{R'(r_{\mathrm{BAO}},t_d(r_{\mathrm{BAO}}))R^2(r_{\mathrm{BAO}},t_d(r_{\mathrm{BAO}}))}\right)^{1/3}.
\end{equation}
Then finally
\begin{equation}
(\Delta\theta^2\Delta z)^{1/3}=\left[\frac{(1+z_{\mathrm{BAO}})\dot{R}'_{\mathrm{BAO}}}{R'(r_{\mathrm{BAO}},t_d(r_{\mathrm{BAO}}))R^2(r_{\mathrm{BAO}},t_d(r_{\mathrm{BAO}}))}\right]^{1/3}\frac{r_d(r_{\mathrm{BAO}})}{1+z_d(r_{\mathrm{BAO}})}.
\end{equation}
See \cite{Biswas:2010xm} for more details. For $\Omega_B$ for the FRW cases and also for $\Omega_{B,\mathrm{out}}$ for the void cases we use the value indicated by Eq.~(\ref{omegab}).

\begin{figure}[htbp]
\centering
\includegraphics[width=10cm]{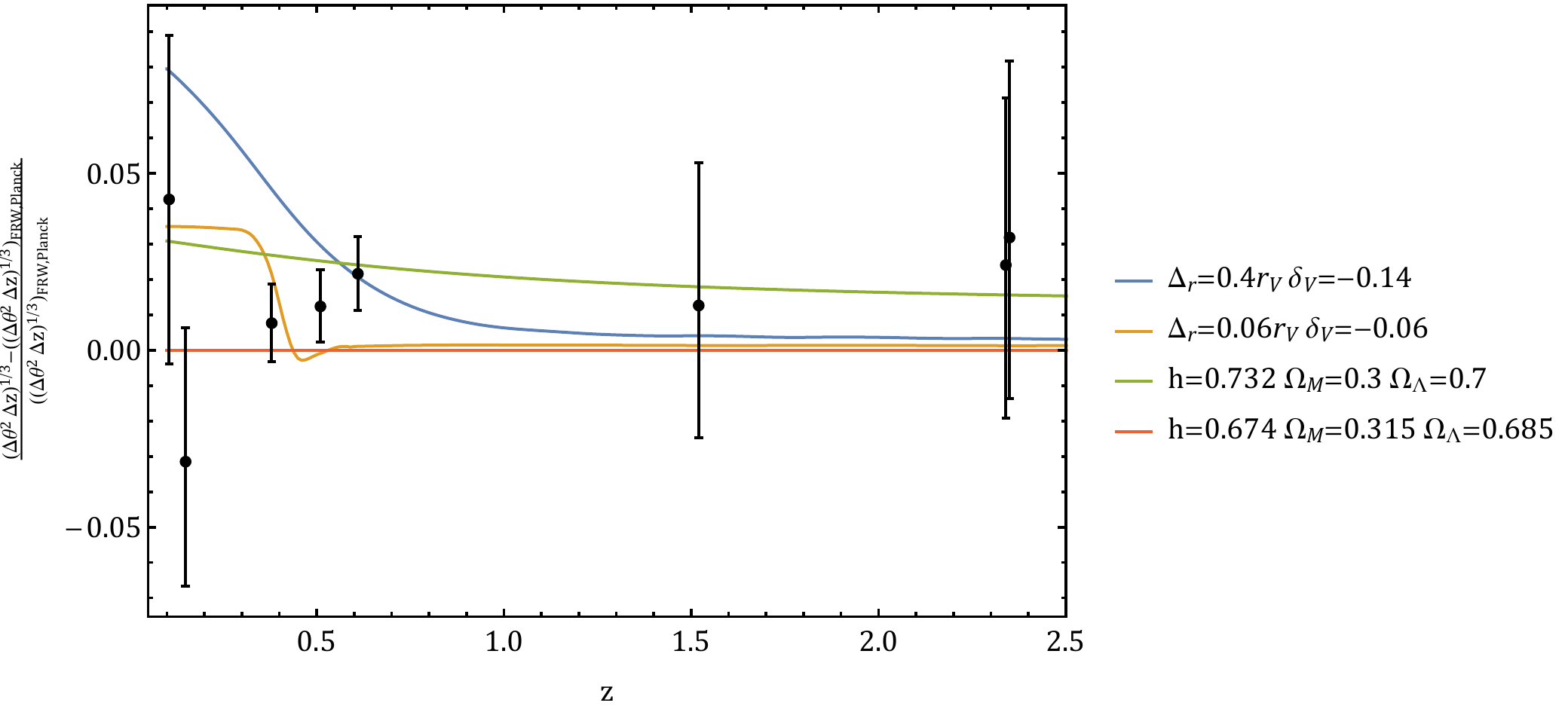}
\caption{Predictions for the BAO observable $(\Delta\theta^2\Delta z)^{1/3}$ and observational data (see text for the details). To make the comparison clearer here the differences from $(\Delta\theta^2\Delta z)^{1/3}$ for the Planck cosmology are plotted.}
\label{bao}
\end{figure}

In Fig.\,\ref{bao} we compare predictions for $(\Delta\theta^2\Delta z)^{1/3}$ for FRW and void cosmologies and measurements by the experiments listed in Table 1 of \cite{Cuceu:2019for}. 
Ref.\,\cite{beutler2011} reports $r_d/D_V^{\mathrm{FRW}}=0.336\pm0.015$ at $z=0.106$, which indicates $(\Delta\theta^2\Delta z)^{1/3}=0.159\pm0.0071$ at $z=0.106$. Similarly from \cite{Ross:2014qpa} $D_V^{\mathrm{FRW}}=(664\pm25) (r_d/r_{d,\mathrm{fid}})\mathrm{Mpc}\,(r_{d,\mathrm{fid}}=148.69\,\mathrm{Mpc})$ at $z=0.15$, which indicates $(\Delta\theta^2 \Delta z)^{1/3}=0.119\pm0.0045$ at z=0.15. From \cite{Alam:2016hwk} $D_V^{\mathrm{FRW}}(r_{d,\mathrm{fid}}/r_d)\mathrm{Mpc}^{-1}=1477\pm 16, 1877\pm19, 2140\pm22\, (r_{d,\mathrm{fid}}=147.78\,\mathrm{Mpc})$ at redshift $z=0.38, 0.51,0.61$, which indicate $(\Delta\theta^2\Delta z)^{1/3}=0.07247\pm0.00079, 0.06290\pm0.00064, 0.05857\pm0.00060$ at redshift 0.38, 0.51 and 0.61. From \cite{Ata:2017dya} $D_V^{\mathrm{FRW}}=(3843\pm147)(r_d/r_{d,\mathrm{fid}})\mathrm{Mpc}\, (r_{d,\mathrm{fid}}=147.78\,\mathrm{Mpc})$ at $z=1.52$, hence $(\Delta\theta^2\Delta z)^{1/3}=0.04421\pm0.00169$ at $z=1.52$. From \cite{Agathe:2019vsu} $D_H/r_d=8.86\pm0.29$ and $D_M/r_d=37.41\pm1.86$ at $z=2.34$ where $D_V^{\mathrm{FRW}}=D_M^{2/3}(zD_H)^{1/3}$, hence $(\Delta\theta^2\Delta z)^{1/3}=0.04320\pm0.0019$ at $z=2.34$. Lastly from \cite{Blomqvist:2019rah} $D_H/r_d=9.20\pm0.36$ and $D_M/r_d=36.3\pm1.8$ at $z=2.35$, hence $(\Delta\theta^2\Delta z)^{1/3}=0.0435\pm0.0020$ at $z=2.35$.

The additional redshift dependence for $(\Delta\theta^2\Delta z)^{1/3}$ of void cosmologies may help fit the data better as was also mentioned in \cite{Biswas:2010xm}, but to draw a more definite and quantitative conclusion we would need a more thorough analysis. 

\section{Limits on a local void from the linear kSZ Effect}
\begin{figure}[H]
\centering
\includegraphics[width=8cm]{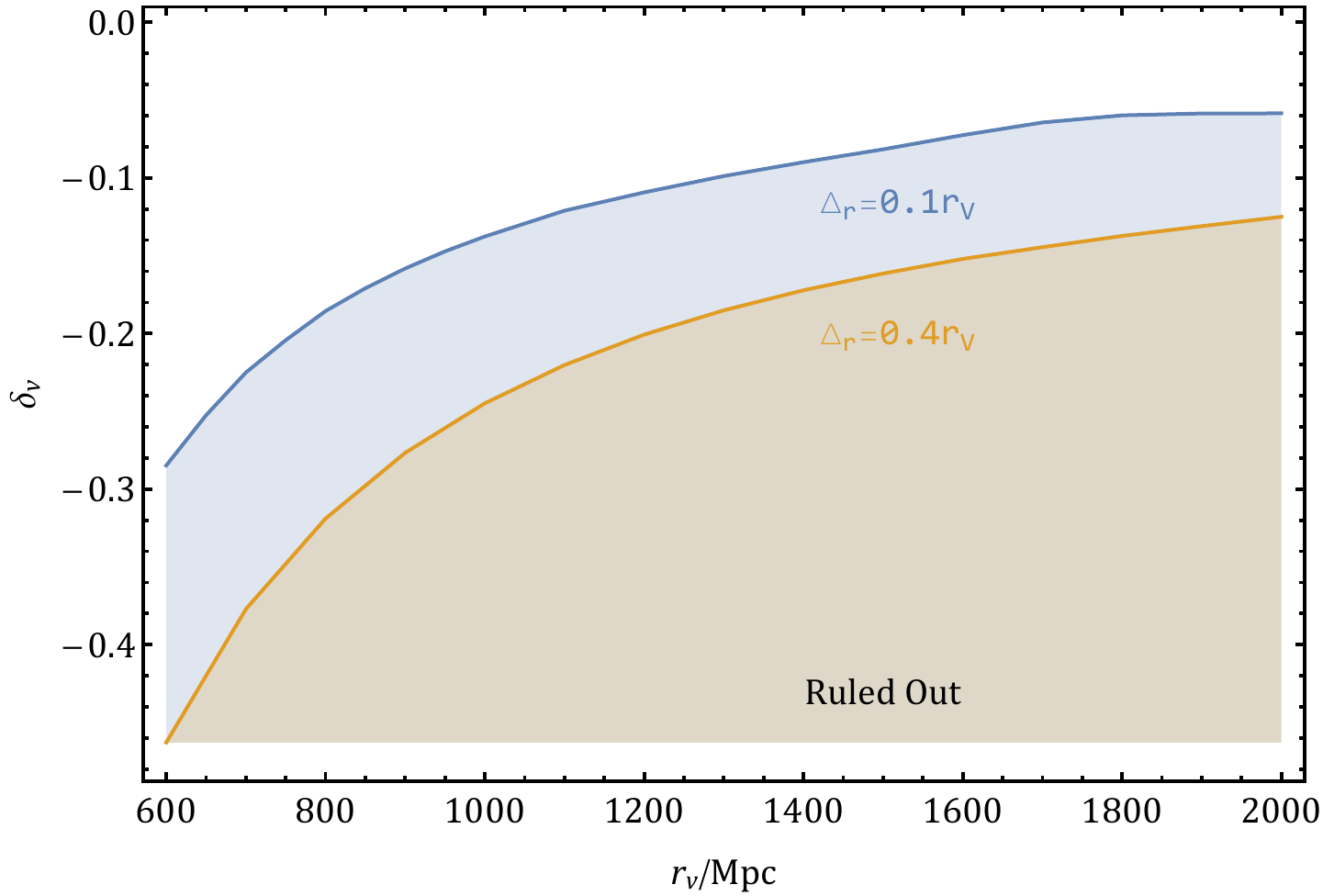}
\caption{The curves show the depth of a local void which corresponds to $T_{\mathrm{CMB}}^2 D_{3000} = 2.9 \mathrm{\mu K}^2$ \cite{George:2014oba,Hoscheit:2018nfl}, and the colored regions are ruled out. 
}
\label{ksz}
\end{figure}

Spatial fluctuations in the electrons in the Universe cause distortions of the CMB spectrum due to interactions between high energy electrons and the CMB photons, which is called kSZ effect \cite{Sunyaev:1980nv}. The temperature perturbation in direction $\hat{n}$ induced by a local void  is given by \cite{Hoscheit:2018nfl}
\begin{align}\label{linearksz}
\Delta T_{\mathrm{kSZ}}(\hat{n}) = T_{\mathrm{CMB}} \int_{0}^{z_e} \delta_{e}(\hat{n},z)\frac{V_H (\hat{n},z) \cdot \hat{n}}{c} d\tau_e~.
\end{align}
Here, $T_{\mathrm{CMB}} = 2.73 \mathrm{K}$, $\delta_e$ is the density contrast of electrons, and $\tau_e$ is the optical depth alonge the line of sight. As in \cite{Zhang:2015}, we choose $z_e = 100$, and we assume
\begin{align}
V_H \simeq [\widetilde{H}(t(z), r(z)) - \widetilde{H}(t(z), r(z_e))] R(t(z), r(z))~,
\end{align}
where, $\widetilde{H} = \dot{R}'/R'$. We use \cite{Zibin:2011, Zibin:2008vk}
\begin{align}\nonumber
\frac{d\tau_e}{dz} &= \sigma_T n_e (z) c \frac{dt}{dz}\\
&= \frac{\sigma_T \theta^2 f_b (1 - Y_{He}/2) \Omega_M (1 + \delta(r(z)))}{24 \pi G m_p} c \frac{dt}{dz}~.
\end{align}
Here, $\sigma_T$ is the Thomson cross section, $f_b$ is the baryon fraction, for which $f_b=0.168$ according to the CMB observations \cite{Aghanim:2018eyx}, $Y_{\mathrm{He}} = 0.24$ is the helium mass fraction, $m_p$ is the proton mass, and $\theta$ is given by
\begin{align}
\theta = (\widetilde{H} + 2 H)~.
\end{align}
We use the Limber approximation \cite{Zibin:2011}:
\begin{align}
C_{\ell} \simeq \frac{16 \pi^2}{(2\ell + 1)^3} \int_0^{z_e} dz \frac{dr}{dz} r(z) F^2(r(z)) P_{\delta}\left(\frac{2\ell + 1}{2 r(z)}, z\right)~.
\end{align}
Here, $C_{\ell}$ is the linear kSZ power at multipole $\ell$, and
\begin{align}
F(r) \equiv \frac{V_H (r)}{c} \frac{d\tau_e}{dz} \frac{dz}{dr}~.
\end{align}
We use the $\Lambda$CDM matter power spectrum $P_{\delta}(k, z)$ from CAMB code \cite{Lewis:1999bs}, where we use cosmological parameters from Planck 2018, which can employ Halofit \cite{Takahashi:2012em} to account for nonlinearities. This simplification would only cause small errors, though, strictly speaking, the matter power spectrum should be calculated for our void cosmology. We define
\begin{align}
D_{\ell} \equiv \frac{\ell (\ell + 1) C_{\ell}}{2 \pi}~.
\end{align}
Then, we make use of the quantity $T_{\mathrm{CMB}}^2 D_{3000}$ to constrain the depth and size of a local void as shown in Fig.~\ref{ksz}. Let us try to understand the dependence of the constraint on $r_V$ as follows. Assuming for simplicity $z$ is small and the Harrison-Zel'dovich spectrum, the contribution to the integrand for $C_l$ from a logarithmic interval in redshift is $z (dr/dz) r(z) V_H(r)^2(d\tau_e/dz)^2(dz/dr)^2P_\delta((2l+1)/[2r(z)],z)\sim z\cdot z \cdot\delta_V^2\cdot z^2\cdot (dt/dz)^2z^{-1}\sim z^3\delta_V^2$, noting $r\sim z$. Only the inside of the void contributes to the integration so the void edge gives the dominant contributions. Hence the constraint may look like $\delta_V>-z_V^{-3/2}\sim -r_V^{-3/2}$. 

\begin{figure}[H]
\centering
\includegraphics[width=8cm]{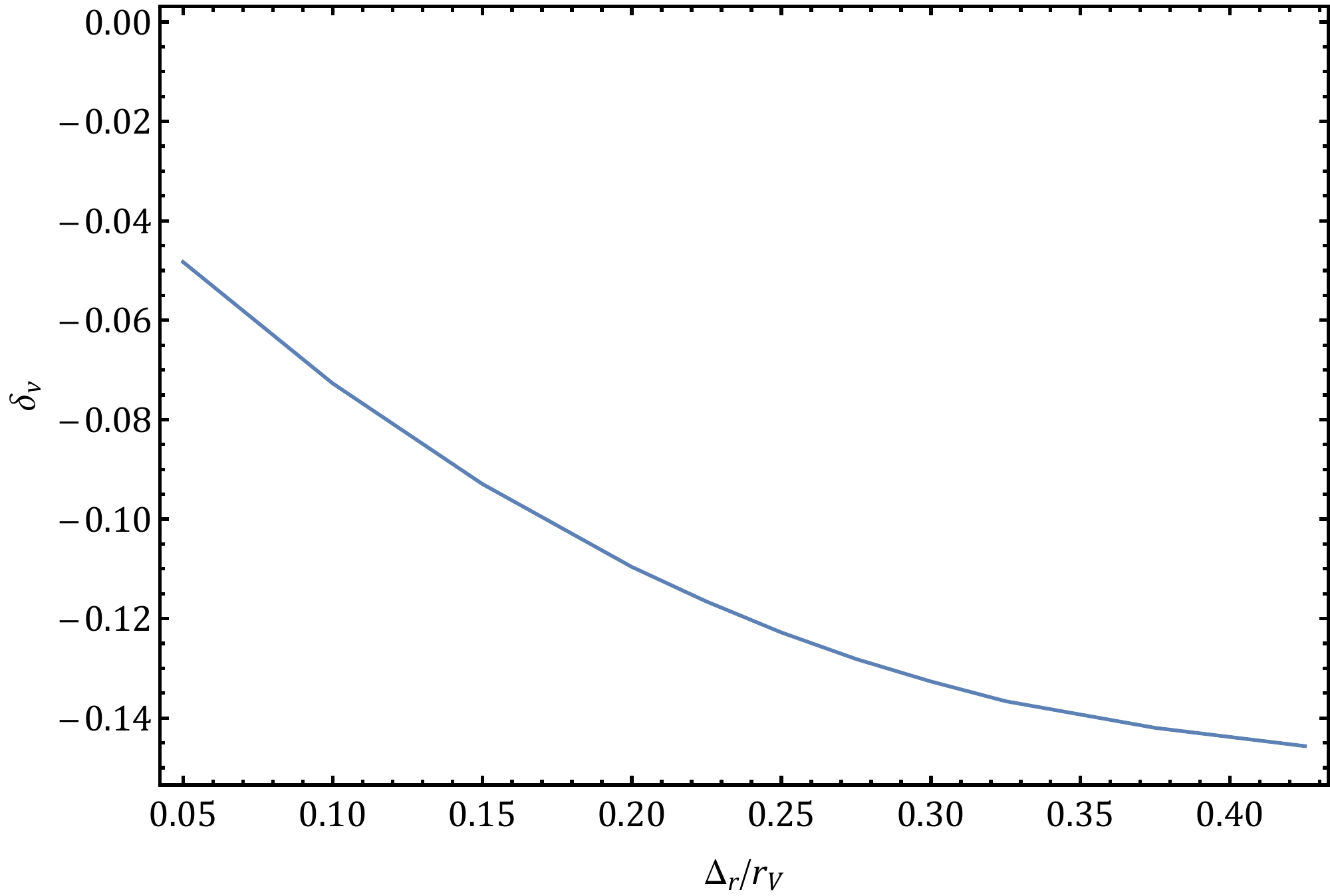}
\caption{The dependence of the kSZ constraint on $\Delta_r$ for $r_V = 1700 \mathrm{Mpc}$.}
\label{widdep}
\end{figure}

We can also find the constrains given by kSZ effect depend on $\Delta_r$, as shown in Fig.~\ref{widdep}, which shows the constraint on the void depth is weaker when $\Delta_r$ is larger. In particular, the great void with the void depth $\delta_V = - 0.14$ can narrowly evade the kSZ constraint for $\Delta_r = 0.4 r_V$. From Fig.~\ref{hubbdep}, the difference in the Hubble parameter at $z=0$ is $2.5\%$ for this case.
\section{Discussion}
We have shown that a gigaparsec-scale local void has the potential to ease the so-called Hubble tension between the local determinations involving SNe and Planck CMB observations. We also discussed implications of such a local void for BAO observations. The depth of such a void and hence to what extent it can ease the tension are restricted by the kSZ effect. Taking this into account, assuming the Planck cosmology for the outside of the void, the internal Hubble parameter can be raised by a few percent due to the presence of the void, thus potentially alleviating the Hubble tension. The presence of such a large void is unexpected in minimal standard cosmological scenarios, but the presence of such a void can be realized by multi-stream inflation. 

Although the mechanism of multi-stream inflation is intuitive and it is known how to relate the early universe model to late time observables, it is important to study multi-stream inflation in more details in the future by, for example, a lattice simulation to study the details of the bubble profile. Also, in the present work we have used a LTB metric with an assumed density profile. It is interesting to understand the density profile from first principles in multi-stream inflation. With these understandings, we would obtain predictions on possible corrections of the distance-redshift relation when the universe is inhomogeneous as a result of multi-stream inflation.

There can be effects other than the kSZ effect of a local void affecting CMB observations. One effect is the difference in the angular diameter distance to the last scattering surface. We have checked that the difference is sufficiently small for the parameter space of a local void we consider, noting that the angular diameter distance in our void cosmology is given by the metric function $R$ in the LTB metric \cite{Biswas:2010xm}. A local void can also induce CMB spectral distortions, but the constraints from this effect wouldn't be as stringent as those from the kSZ effect \cite{Caldwell:2007yu}. A local void in principle should also affect the CMB observations through the integrated Sachs-Wolfe effect. But quantifying this effect precisely in void cosmology appears challenging partly because cosmological perturbation on an inhomogeneous background can be very complicated \cite{Biswas:2010xm,Moss:2010jx}. This would be particularly so for voids with a thick edge, whereas for voids with a sharp edge we may find approximate methods by treating the inside and outside different FRW universes. 

In this work we have assumed that we locate at the center of the void. In general, we should be located away from the exact center of a local void, and there would be additional observational effects associated with this \cite{Blomqvist:2009ps,Nistane:2019yzd}. The observations would also be affected from deviations from spherical symmetry of a local void. These two effects would determine the amount of fine-tuning associated with our scenario. 

A few more comments are in order. It would also be interesting and important to investigate how the parameter determinations of \cite{Bonvin:2016crt} are affected by the presence of a local void. Our scenario would be tested by observations of high-redshift SNe or galaxies, the redshift distribution of transients such as gamma-ray or gravitational-wave bursts, and future CMB experiments due to the kSZ effect. 

\acknowledgments
	 
We thank Xingang Chen and Zhong-Zhi Xianyu for delightful discussions. Q. D. would like to thank Xingwei Tang for her kindly help in code optimization and computational setup.

\end{document}